\documentclass[twocolumn,showpacs,preprintnumbers,amsmath,amssymb]{revtex4}

\usepackage{graphicx}
\usepackage{dcolumn}
\usepackage{bm}
\begin{document}
\preprint{}


%
%

\title{Looking for non-perturbative effects of the electroweak interaction
\\at the LHC}

\author{Boris A. Arbuzov }
\affiliation{Lomonosov Moscow State University\\ Leninskie gory 1, 119991 Moscow, Russia}%
\email{arbuzov@theory.sinp.msu.ru}
\author{Ivan V. Zaitsev}
\affiliation{Lomonosov
Moscow State University\\ Leninskie gory 1,119991 Moscow, Russia}
\date{\today}


\begin{abstract}
We are considering a possibility for detecting non-perturbative effect
in process of top pair production in association with a high $p_T$ photon.
Starting from previous results on two solutions for a spontaneous generation of wouldbe effective interaction of a top pair with
a pair of electro-weak bosons, we show, that
 a solution with effective cut-off $\Lambda \simeq 10^2\,TeV$ is already contradicting to existing data, while the other one with $\Lambda \simeq 10^{16}\,TeV$ (just the Planck scale) could be reconciled with data and give predictions for process $p+p\to\bar t t \gamma+X$, which could be effectively checked at the LHC with $\sqrt{s}=13\,TeV$. The confirmation of the predictions would mean a strong support for the existence of non-perturbative effects in the
 electro-weak interaction.

 \end{abstract}

\pacs{12.15.-y, 12.15.Ji, 14.70.Fm, 14.80.Ec}
\keywords{anomalous triple boson interaction; W-ball;
spontaneous generation of an effective interaction}

\maketitle
\section{Introduction}

The two  constituents of the Standard Model, QCD and the
EW theory are renormalizable field theories in which a perturbation theory
can be duly developed. In the region of the applicability of this theory
the totality of data shows excellent agreement with the theory, that substantiates
an assurance in an adequate description of the elementary particles physics in the framework
of the SM. However, at least, in the strong interaction physics, which is described by QCD,
the presence of the non-perturbative contributions is doubtless. This contributions manifest
themselves in a small momenta area. For example such effects arise as nonzero vacuum averages: the gluon condensate $V_2 \equiv <\frac{\alpha_s}{\pi}G_{\mu
\nu}^aG_{\mu \nu}^a>$ and the quark condensate $< \bar q\,q >$. The behaviour of strong running coupling $\alpha_s(Q^2)$ itself definitely
needs non-perturbative modification, because the perturbative expression contains
a singularity at $Q^2=\Lambda_{QCD}^2$.

There are methods of dealing with the
non-perturbative effects, {\it e.g.} a description of the effects
in the lattice approach~\cite{Kogut} and the effective interaction
approach~\cite{Weinberg}. Here we would rely on the second option,
which was started by the famous Nambu - Jona-Lasinio model~\cite{NJL1,NJL2}.
With application to quark structure of
hadrons this model adequately describes the low momenta region,
see {\it e.g.} review \cite{ERV}.

In dealing with method of effective interactions the most important problem consists
in their origin in the framework of the initial renormalizable gauge theory.
It comes out, that promising prospects here has the conception of a spontaneous
generation of effective interactions, which is based on methods being developed in
the theory of the superconductivity. We rely here on the compensation approach by
N.N. Bogoliubov~\cite{Bog1}, which was successfully applied just to the superconductivity.
In application to effective interactions of the Standard Model the approach may be
also used (see details in book~\cite{ABOOK}).
Indeed, this approach was
applied to spontaneous generation of the Nambu--Jona-Lasinio
interaction in work \cite{AVZ06} and of three-gluon interaction~(\ref{3gluon}) in
work \cite{AZ13}. In the course of this application, form-factors
inherent to corresponding interactions are uniquely defined
\footnote{Of course, in the framework of an approximation.}. As an
additional confirmation of applicability of the method to
non-perturbative quantities, value $V_2$ of the gluon condensate was
calculated \cite{AZ13} in the agreement with its phenomenological
value. Emphasize, that the structure of a variety  of solutions in QCD indicates
that similar effects may be present in the electroweak theory as well.

In the course of considering options of a spontaneous generation of an
effective interaction, which acts in a limited region of the momentum space
we proceed in the way, which we illustrate with an example of the
three-gluon interaction of the form
\begin{equation}
L_{eff}^{QCD}\,=\,-\frac{G}{3!}\cdot F\cdot f_{abc}G^a_{\mu\nu}G^b_{\nu\rho}G^c_{\rho\mu};
\label{3gluon}
\end{equation}
where $G^a_{\mu\nu}$ is a gauge covariant gluon field, $G$ is a dimensional coupling parameter, $f_{abc}$
are structure constants of the color $SU(3)$ and symbol $F\equiv F(p1,p2,p3)$ corresponds to
a form-factor, which guarantees the action of interaction (\ref{3gluon}) just in
the limited region of small momenta $p_i$. This wouldbe low momenta effective interaction was earlier considered in the framework of phenomenological models for low momenta QCD interaction~\cite{kb,aab}.
According to the compensation approach form-factor $F$ is a solution of a compensation equation, which is formulated according to the procedure
{\it add-subtract}~\cite{Bog1}.
Indeed, let us consider QCD Lagrangian consisting, as usually, of free part $L_0$ and of
interaction part $L_{int}$
\begin{equation}
L_{QCD}\,=\,L_0\,+\,L_{int}.\label{L0i}
\end{equation}
Then we take expression~(\ref{3gluon}) and redefine expression~(\ref{L0i}) in the
following way
\begin{eqnarray}
& &L_{QCD}\,=\bar L_0\,+\,\bar L_{int}; \nonumber\\
& &\bar L_0\,=\,L_0\,-\,L_{eff}^{QCD}; \label{LN}\\
& &\bar L_{int}\,=\,L_{int}\,+\,L_{eff}^{QCD}. \nonumber
\end{eqnarray}
Then we have to define the compensation equation for the effective interaction. For the purpose
we impose the condition,
that for the field theory with Lagrangian $\bar L_0$ (\ref{LN})
 the 3-gluon vertex with Lorentz structure of interaction~(\ref{3gluon}) has
to vanish, i.e. total contribution of the vertex into the free Green function has to be equal to zero.
In the same time contribution into interaction Lagrangian will not vanish by the reason of different sign before the corresponding term.

The compensation equation evidently has the trivial solution $L_{eff}^{QCD}\equiv0$ and for the most cases of wouldbe effective Lagrangians it is just the only solution. However we have succeeded to show, that here the compensation equation
has also two non-trivial solutions. The first one has a small effective cut-off and a large value for
gauge coupling. This solution can be related just to QCD~\cite{AZ13}. The second solution, on the contrary, corresponds to a large effective cut-off and the relatively small gauge coupling, that
may be related to the electro-weak interaction~\cite{AZ12}.

In the framework of the approach the generation of the effective interaction
 being
analogous to effective interaction~(\ref{3gluon}) is also quite natural
 \begin{eqnarray}
& &L_{eff}\,=\,-\frac{G_W}{3!}\cdot F_{W}\cdot\epsilon_{abc}W^a_{\mu\nu}W^b_{\nu\rho}W^c_{\rho\mu};\label{3boson}\\
& &G_W=\frac{g\, \lambda}{M_W^2};\; W_{\mu \nu}^3\,=\,\cos\theta_W\,Z_{\mu\nu}\,+\,\sin\theta_W\,A_{\mu\nu};
\nonumber
\end{eqnarray}
where $g$ is the electro-weak gauge coupling, indices $a,b,c$ take
three values $\overline {1,3}$ and the third boson $W_{\mu \nu}^3$ is
the well-known composition of neutral vector bosons $Z$ and $\gamma$.
The wouldbe existence of interaction~(\ref{3boson}) was discussed in works
~\cite{Hag1,Hag2}. Again there is form-factor $F_W$, which is to be obtained as a solution
of a compensation equation. The first attempt to consider the spontaneous
generation of the interaction was made in work~\cite{BA92}.
Interaction~(\ref{3boson}) would lead to effects {\it e.g.} in the pair
electro-weak bosons production and was studied in experiments. The
best limitation for parameter $\lambda$ is provided by recent data of
CMS Collaboration~\cite{CMSlam}
\begin{equation}
-0.011<\lambda<0.011.\label{CMSlam}
\end{equation}
In the electro-weak interaction necessity of non-perturbative
contributions nowadays is not so evident as in QCD. However the
structure of gauge theories is the same for both cases. One might
expect similar features in both three-boson interactions~(\ref{3gluon}) and (\ref{3boson}).
Both the Nambu--Jona-Lasinio interaction and
interaction~(\ref{3gluon}) are supposed to act in a low momenta region.
This means, that in both cases form-factors are present, which
guarantee decreasing of intensity of the interactions for large
momenta. In the original NJL interaction \cite{NJL1,NJL2} a cut-off
was introduced for the purpose.

\section{A quest for non-perturbative effects of the electroweak interaction.}

The direct test for non-perturbative effects of interaction~(\ref{3boson}) consist in an
experimental study of the pair weak boson production and for the moment it leads to
limitation~(\ref{CMSlam}). However there are also other effects, which might be even more promising. In work~\cite{AZ17} we have studied contributions of interaction~(\ref{3boson}) to process of the Higgs boson production in association with top quark pair and have shown, that current table value of ratio $\mu$~\footnote{Throughout the article $\mu$ means, as usually, a ratio of an actual value and its SM value.}    for the process $\mu = 2.3^{+0.7}_{-0.6}$ \cite{PDG} leads to estimate $\lambda = -0.0057^{+0.0028}_{-0.0039}$, that evidently is consistent with zero, but it allows also an existence of a wouldbe significant non-perturbative contribution. The last data for this effect at $\sqrt{s} =13\,TeV$ being obtained by CMS~\cite{muCMS13} and ATLAS~\cite{muATLAS13}, are correspondingly
\begin{equation}
\mu = 1.26^{+0.31}_{-0.26};\quad
\mu = 1.32^{+0.28}_{-0.26};\label{mu13ATLAS}
\end{equation}
with average $\mu=1.29^{+0.23}_{-0.20}$. According to~\cite{AZ17} this correspond to value $\lambda=  -0.0013\pm 0.0010$. Both estimates for $\lambda$ are consistent with zero, but we may also note, that the PDG value and
estimates~(\ref{mu13ATLAS})  do not contradict {\it e.g.} to
\begin{equation}
\lambda \simeq -0.003. \label{Laverage}
\end{equation}
The question is, if we may hope to
study possible non-perturbative effects on the level of value~(\ref{Laverage})? In any case
it seems advisable to look for other ways for studying of non-perturbative effects.

In particular, we may rely on the possible non-perturbative effect, which was discussed in work~\cite{AZ16}.
In this work the spontaneous generation of the would-be four-boson interaction of the form,
proposed in work~\cite{BB1}, is considered.
Let us  consider a possibility of a spontaneous
generation of the following
effective interaction of electroweak gauge bosons~\cite{AZ16}
\begin{eqnarray}
& & L_{eff}^{4W} =
-\,\frac{G_2}{8} W^a_\mu W^a_\mu W^b_{\rho \sigma}
W^b_{\rho \sigma} - \frac{G_3}{8}\,W^a_\mu W^a_\mu B_{\rho \sigma}
B_{\rho \sigma} -\nonumber\\
& &\frac{G_4}{8}\,Z_\mu Z_\mu W^b_{\rho \sigma} W^b_{\rho \sigma}
- \frac{G_5}{8}\,Z_\mu Z_\mu B_{\rho \sigma} B_{\rho \sigma} .
\label{eq:LeffWZ}
\end{eqnarray}
where we maintain the gauge invariance for the electromagnetic
field. Here indices $a$ correspond to charged $W$-s, that is they
take
values $1,\,2$, while an index $b$
corresponds to three components of $W$ defined by the initial
formulation
of the electro-weak interaction. While considering solutions of compensation equations, corresponding to wouldbe interaction~(\ref{eq:LeffWZ}) we have to introduce an effective cut-off $\Lambda$, which is also a parameter to be defined in case of existence of a non-trivial solution of the set of equations~\cite{AZ16}.

\begin{figure}
\begin{center}
\includegraphics[scale=0.40]{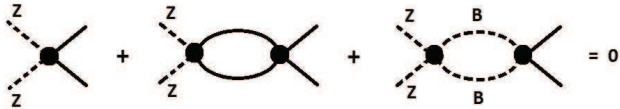}
\caption {Diagram representation of an example of compensation equations. Here simple line represent $W$-s,
dotted lines represent $B$ or  $Z$ -bosons according to letter notices.}\label{fig:ExEq}
\end{center}
\end{figure}

We consider the set of compensation equations corresponding to wouldbe interaction~(\ref{eq:LeffWZ}) in the one loop approximation.  In doing this we introduce a formfactor being proportional to a step function, defining effective cut-off $\Lambda$,
\begin{equation}
F(p_i)\,\simeq\,\theta(\Lambda^2 - (p1+p2)^2);\label{FF}
\end{equation}
where $p_{1,2}$ are Euclid momenta of virtual bosons at FIG.~\ref{fig:ExEq}.
$\Lambda$ is also a parameter to be defined in case of an existence of a non-trivial solution of the set of compensation equations~\cite{AZ16}. In the case of Nambu - Jona-Lazinio theory we have consider the one-loop approximation as well as two loop one, and found that the difference in resulting parameters is of order of magnitude of $10$  percent. We would use this result for an estimation of the wouldbe accuracy.

An example of these equations is presented in diagram form at FIG.~\ref{fig:ExEq}. Full set determines couplings for each vertex as well as the
cut-off.
There are two solutions which
satisfy our conditions. The procedure is described in details in work~\cite{AZ16}. Namely, for the both solutions $G_3\,=\,G_5\,=\,0$ and
\begin{eqnarray}
& &G_2 = -2.20\cdot 10^{-33} TeV^{-2};\,G_4 = -4.96\cdot 10^{-33} TeV^{-2}\nonumber\\
& &\Lambda = 8.69\cdot10^{19} GeV;\label{solutionhighG}\\
& &G_2 = -6.24\cdot 10^{-5} TeV^{-2};\;G_4 = -9.38\cdot 10^{-5} TeV^{-2}\,\nonumber\\
& &\Lambda = 5.23\cdot 10^{5}\,GeV. \label{solutionlowG}
\end{eqnarray}
\begin{figure}
\begin{center}
\includegraphics[scale=0.32]{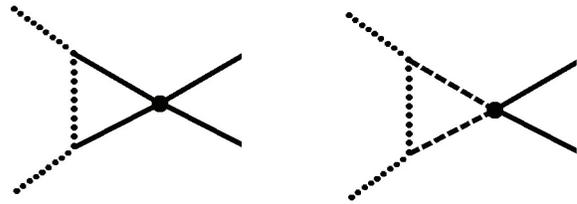}
\caption{Diagram representation of $\bar t t W W$ vertex. Here left-side lines represent $t$-quarks,
dotted lines represent $Z$-bosons and
full lines correspond to $W^\pm$. Right-side vertices correspond to a solution for
interaction~(\ref{eq:LeffWZ}) }
\label{fig:WWTt}
\end{center}
\end{figure}

Solutions~(\ref{solutionhighG}, \ref{solutionlowG}) lead to definition of effective interaction
with participation of the $t$-quarks in accordance to diagrams in FIG.~\ref{fig:WWTt}.
\begin{equation}
\frac{G_{W \bar t t}}{4}\bar t t W^b_{\mu \nu}\,W^b_{\mu \nu};\label{ttw}
\end{equation}
were $b$, just as above,  takes 3 values. Solutions~(\ref{solutionhighG}, \ref{solutionlowG}) provide  the following values of parameters $G_{W \bar t t}$ in~(\ref{ttw}) for interaction~(\ref{eq:LeffWZ})
\begin{equation}
G_{W \bar t t}^{(1)} = 1.50\cdot 10^{-8} GeV^{-3},\; \Lambda =8.69\cdot10^{19}\,GeV;\label{2sol2}
\end{equation}
\begin{equation}
G_{W \bar t t}^{(2)} = 4.25\cdot 10^{-8} GeV^{-3},\, \Lambda =5.23\cdot10^{5}\,GeV.
\label{2sol1}
\end{equation}
Interaction~(\ref{ttw}) leads to effects in associated top pair production with electroweak
bosons $W^\pm,\,Z$, which for the first solution are presented in work~\cite{AZ16}. The new
data for the process at $\sqrt{s}=13\,TeV$~\cite{ttw13} results in contradiction with the second solution~(\ref{2sol1}), while is consistent with the first  solution~(\ref{2sol2}) as well as with zero value for
$G_{W \bar t t}$. Let us show in TABLE~\ref{tab:table1} results for cross sections of the relevant processes at $\sqrt{s}=13\,TeV$ and corresponding ratios $\mu$ for  solutions~(\ref{2sol2}) and~(\ref{2sol1}).
We immediately see, that values $\mu(G_{W \bar t t}^{(2)})$ at $\sqrt{s}=13\,TeV$ for process \begin{equation}
p\,+\,p\,\to\,\bar t\,t\,W^\pm\,+\,X;\label{TTW1}
\end{equation}
contradict result of recent
work~\cite{ttw13}, which presents the following data for parameters under discussion
\begin{equation}
\mu(W \bar t t)=1.23^{+0.19}_{-0.18}(stat)^{+0.20}_{-0.18}(syst)^{+0.13}_{-0.12} (theo) \,;\qquad\nonumber
\end{equation}
\begin{equation}
\mu(Z\, \bar t t)=1.17^{+0.11}_{-0.10}(stat)^{+0.14}_{-0.12}(syst)^{+0.11}_{-0.12}(theo) \,.
\label{exp}
\end{equation}
Experimental result~(\ref{exp}) evidently excludes the second solution~(\ref{2sol1}). However the first solution~(\ref{2sol2}) has no contradiction, as well as the perturbative SM prediction $\mu=1$.

We would emphasize, that the first solution corresponds to very high
effective cut-off, which can be related to the Planck mass $M_{Pl}=1.221\cdot10^{19}\,GeV/c^2$.
Indeed, prediction~(\ref{2sol2}) is close to the following quantity
\begin{equation}
\Lambda^{(1)} \simeq 2\pi M_{Pl}\,c^2 = 7.67\cdot 10^{19}\,GeV.\label{MPl}
\end{equation}
In case of a real occurrence of solution~(\ref{solutionhighG}, \ref{2sol2}), we may hope for better understanding of  the well-known problem of the hierarchy~\cite{Witten}. Indeed, the approach under the consideration is based on the electroweak scale parameters and the Planck mass scale arises here as a result of the spontaneous generation of the corresponding effective interaction. That is the Planck scale and the habitual electroweak scale may become mutually consistent.
\bigskip
\begin{center}
\begin{table}
\caption{\label{tab:table1}SM results for cross-sections of processes $p+p\to\bar t t V$ at
$\sqrt{s}=13\, TeV$ and predictions for
ratios $\mu$ being due to effective interaction~(\ref{ttw}).}
\begin{ruledtabular}
\begin{tabular}{l c c c}
\hline
channel & $\sigma_{SM}\,pb,\,13\,TeV  $ & $\mu(G_{W \bar t t}^{(1)})$ & $\mu(G_{W \bar t t}^{(2)})$
 \\
\hline
$\bar t t W$ & $0.77^{+0.18}_{-0.16}$  & $1.27$ & $3.18$ \\
\hline
$\bar t t Z$ & $0.99^{+0.15}_{-0.13}$ & $1.06$ & $1.50$ \\
\hline
\end{tabular}
\end{ruledtabular}
\end{table}
\end{center}
Let us draw attention to process~(\ref{TTW1}). The dependence on a transverse momentum of the $W$ boson allows to extract the effect of interaction~(\ref{ttw}). In TABLE~{\ref{tab:table2}} we show this dependence for three value of $G_{W \bar t t}$ one of which is just value~(\ref{2sol2}) and the two others are $10\%$
up and down. We do take this variation of the coupling constant due to estimate of the accuracy of the approach being by the order of magnitude
$10-15\%$~\cite{Arb1}.
Note, that the difference between values~(\ref{2sol2}) and (\ref{MPl}) is also of the same order of magnitude.
\begin{center}
\begin{table}
\caption{\label{tab:table2}SM LO results for cross-sections of processes $p+p\to\bar t t W^\pm$ at
$\sqrt{s}=13\, TeV$ and predictions for three values
$\mu_i$ for intervals of $p_T$ of the $W$ boson in correspondence to values of wouldbe coupling
$G^i_{W \bar t t}$:\,\, $G_{W \bar t t}^1\, =\, 1.35\cdot10^{-8}\,GeV^{-3},G_{W \bar t t}^2 =1.5\cdot10^{-8}\,GeV^{-3},\,G_{W \bar t t}^3
 =1.65\cdot10^{-8}\,GeV^{-3}$.}
\begin{ruledtabular}
\begin{tabular}{l c c c c }
\hline
$p_T(W)\,GeV$ & $\sigma_{SM}\,fb $ & $\mu_1$& $\mu_2$
&
$\mu_3 $   \\
\hline
$>100$ & $ 223.29$ & $1.51$ & $1.63$ &
$ 1.76$\\
\hline
$>150$ & $123.60$  & $1.90$ & $2.11$ & $2.35$\\
\hline
$>200$ & $72.82$  & $2.46$ & $2.81$ & $3.19$\\
\hline
$>250$ & $ 44.58$ & $3.28$ & $3.81$ &
$ 4.41$\\
\hline
$>300$ & $ 28.40$ & $4.39$ & $5.18$ &
$ 6.07$\\
\hline
\end{tabular}
\end{ruledtabular}
\end{table}
\end{center}
With actual integral luminosity $L \simeq 30\,fb^{-1}$ we
see, that optimal cut-of may be $p_T > 200\,GeV$. We have considered also possibility of registration of the effect in process $p+p\to\bar t t \gamma$. The results for the same values of coupling $G_{W \bar t t}$ are shown in TABLE~{\ref{tab:table3}}. Of course, the values for the cross-sections are smaller, than for the process with $W$ bosons, however, preferences in the registration efficiency of photons may make the study of the process promising. For example
with $p_T > 450\,GeV$ the effect in $\mu$ is significant, namely for integral luminosity $L = 35\,fb^{-1}$ there may around $600$ events of the process.
\begin{center}
\begin{table}
\caption{\label{tab:table3}SM LO results for cross-sections of processes $p+p\to\bar t t \gamma$ at
$\sqrt{s}=13\, TeV$ and predictions for three values
$\mu_i$ in dependence of minimal value of $p_T(\gamma)$ in correspondence to values of wouldbe coupling
$G^i_{W \bar t t}$, the same as in TABLE~\ref{tab:table2}.}
\begin{ruledtabular}
\begin{tabular}{l c c c c }
\hline
$p_T(\gamma)\,GeV$ & $\sigma_{SM}\,fb $ & $\mu_1$& $\mu_2$
&
$\mu_3 $   \\
\hline
\hline
$ >300 $ & $34.61$  & $1.30$ & $1.37$ & $1.45$ \\
\hline
$ >350 $ & $20.62 $  & $1.47$ & $1.58$ & $1.70$ \\
\hline
$>400$ & $ 12.64$ & $1.72$ & $1.88$ &
$ 2.07$
\\
\hline
$ >450 $ & $ 7.97 $ & $2.05$ & $2.30$ &
$ 2.57$
\\
\hline
$>500$ & $5.16$ & $2.50$ & $2.85$ &
$3.24$\\
\hline
$>600$ & $2.29$ & $3.84$ & $4.51$ &
$5.25$\\
\hline
\end{tabular}
\end{ruledtabular}
\end{table}
\end{center}
Data for $\sqrt{s}=7\,TeV$ and integral luminosity $4.59\,fb^{-1}$ \cite{ATLAS7} agree the SM prediction. For illustration we show in TABLE~{\ref{tab:table4}}
our corresponding results. The results show no contradiction with experiment \cite{ATLAS7}, which has used cut $E_T(\gamma) > 20\,GeV$.
\begin{center}
\begin{table}
\caption{\label{tab:table4}SM LO results for cross-sections of processes $p+p\to\bar t t \gamma$ at
$\sqrt{s}=7\, TeV$ and predictions for three values
$\mu_i$ in dependence on minimal value of $p_T(\gamma)$ in correspondence to values of wouldbe coupling
$G^i_{W \bar t t}$, the same as in TABLE~\ref{tab:table2}}
\begin{ruledtabular}
\begin{tabular}{l c c c c }
\hline
$p_T(\gamma)\,GeV$ & $\sigma_{SM}\,fb $ & $\mu_1$& $\mu_2$
&
$\mu_3 $   \\
\hline
\hline
$ >20 $ & $458.8$  & $1.002$ & $1.0024$ & $1.003$ \\
\hline
$ >50 $ & $190.1$  & $1.0044$ & $1.0048$ & $1.0064$ \\
\hline
$ >100 $ & $70.1$  & $1.011$ & $1.012$ & $1.015$ \\
\hline
$ >200 $ & $14.6 $  & $1.04$ & $1.05$ & $1.06$ \\
\hline
$>300$ & $ 3.84$ & $1.11$ & $1.14$ &
$ 1.17$
\\
\hline
\end{tabular}
\end{ruledtabular}
\end{table}
\end{center}

\section{Conclusions}
The problem of wouldbe non-perturbative contributions of the electroweak interaction
is a fundamental one. In the present article we discuss possibilities for a registration of
non-perturbative effects in processes $p+p\to \bar t t W^\pm$ and $p+p\to \bar t t \gamma$.
The predicted effects seem to be achievable for LHC experiments at $\sqrt{s}=13\,TeV$ with
the actual luminosity. A confirmation of our predictions would prove the existence of the non-perturbative contributions. We have already emphasized that the predictions under discussion
correspond to effective cut-off $\Lambda$~(\ref{MPl}) being close to the Planck boundary, that may open a way to understanding of the famous hierarchy problem.


\begin{thebibliography}{**}
\bibitem{Kogut} J. B. Kogut, A review of the lattice theory aproach to Quantum Chromodynamics, Rev. Mod. Phys., 55 (1983), p. 775.
\bibitem{Weinberg} S. Weinberg, {\it The Quantum Theory of Fields, Vol 1: Foundations}, Chap. 12,
Cambridge University Press, Cambridge, 1995.
\bibitem{NJL1} Y. Nambu and G. Jona-Lasinio,
Dynamical model of elementary particles based on an analogy with superconductivity, I,
Phys. Rev., 122 (1961), p. 345.
\bibitem{NJL2} Y. Nambu and G. Jona-Lasinio, Dynamical model of elementary particles based
on an analogy with superconductivity, II, Phys. Rev., 124 (1961), p. 246.
\bibitem{ERV} D. Ebert, H. Reinhardt and M. K. Volkov, Effective hadron theory of QCD, Prog. Part. Nucl. Phys., 33 (1994), p. 1.
\bibitem{Bog1} N. N. Bogoliubov, On compensation principle and
selfconsistent field method, Soviet Phys.-Uspekhi, 67 (1959), p. 236.
\bibitem{ABOOK} B. A. Arbuzov, {\it Non-perturbative Effective
    Interactions in the Standard Model}, De Gruyter, Berlin, 2014.
\bibitem{AVZ06}B. A. Arbuzov, M. K. Volkov and I. V. Zaitsev, NJL model derived from QCD, Int. J. Mod. Phys. A, 21 (2006), p. 5721.
\bibitem{AZ13} B. A. Arbuzov and I. V. Zaitsev, On a possibility to elimate the Landau
pole in QCD, Int. J. Mod. Phys. A, 28 (2013), Article 1350127.
\bibitem{kb} S. K. Kim and M. Baker, Consequences of gauge invariance for the interacting vertices in nonabelian gauge theories, Nucl. Phys. B, 164 (1980), p. 152.
\bibitem{aab} A. I. Alekseev, B. A. Arbuzov and  V. A. Baikov, Infrared asymptotics of
gluon Green's functions in Quantum Chromodynamics,
Theor. Math. Phys., 52 (1982), p. 739.
\bibitem{AZ12} B. A. Arbuzov and I. V. Zaitsev, LHC would-be excess as a non-perturbative effect of the electro-weak interaction,  	
 Phys. Rev. D, 85 (2012), Article 093001.
\bibitem{Hag1} K. Hagiwara, R. D. Peccei, D. Zeppenfeld and K.
    Hikasa, Probing the weak boson sector in $e^+ e^- \to W^+ W^-$, Nucl. Phys. B, 282 (1987), p. 253.
\bibitem{Hag2} K. Hagiwara, S. Ishihara, R. Szalapski and D.
    Zeppenfeld, Low-Energy effects of new interactions in the electroweak boson sector, Phys. Rev. D, 48 (1993), p. 2182.
\bibitem{BA92} B. A. Arbuzov, On a dynamical breaking of the electroweak symmetry, Phys. Lett. B, 275 (1992), p. 179.
\bibitem{CMSlam} A. M. Sirunyan {\it al.} (CMS Collaboration), Search for anomalous couplings in boosted $WW/WZ \to l \nu q \bar q$ in proton-proton collisions at $\sqrt{s} = 8\, TeV$, Phys. Lett. B, 772 (2017), p. 21.
\bibitem{AZ17} B. A. Arbuzov and I. V. Zaitsev, Associated heavy quark pair production with Higgs as a tool for a search for non-perturbative effects of the electroweak interaction at the LHC, Phys. Lett. B, 772 (2017), p. 184.
\bibitem{PDG} M. Tanabashi {\it et al.} (Particle Data Group),
2018 Review of particle physics, Phys. Rev. D, 98 (2018), Article 030001.
\bibitem{muCMS13} A. M. Sirunyan {\it al.} (CMS Collaboration), Observation of $t \bar t H$
production, Phys. Rev. Lett., 120 (2018), Article 231801.
\bibitem{muATLAS13} M. Aabud {\it et al.} (ATLAS Collaboration), Observation of Higgs boson production in association with top quark pair at the LHC with the ATLAS detector, Phys. Lett. B, 784 (2018), p. 173.
\bibitem{AZ16} B. A. Arbuzov and I. V. Zaitsev, On a possible effective four-boson interaction and its implications at the  upgraded LHC, PTEP, 2016 (2016),  093B02.
\bibitem{BB1} G. Belanger and F. Boudjema, Probing quartic couplings of weak bosons through three vectors production at a 500-GeV NLC, Phys. Lett. B, 288 (1992),
 p.  201.
\bibitem{ttw13} A. M. Sirunyan {\it et al.} (CMS Collaboration), Measurement of the cross section for top quark pair production in association with a W or Z boson in proton-proton collisions at $\sqrt{s} = 13\, TeV$, JHEP, 08 (2018), p. 011.
\bibitem{Witten} E. Witten, Dynamical breaking of supersymmetry, Nucl. Phys. B, 188 (1981),  p.    513.
\bibitem{Arb1} B. A. Arbuzov, Spontaneous generation of effective interaction in a renormalizable quantum field theory model, Theor. Math. Phys., 140 (2004), p. 1205.
\bibitem{ATLAS7} G. Aad {\it et al.} (ATLAS Collaboration), Observation of top-quark pair production in association with a photon and measurement of the $ t \bar t \gamma$ production cross section in $p\,p$ collisions
    at $\sqrt{s} = 7\,TeV$ using the ATLAS detector, Phys. Rev. D, 91 (2015), Article 072007.
\end{thebibliography}
\end{document}